\documentclass[useAMS]{mn2e}
\usepackage{times}
\usepackage{graphics}
\usepackage{color}
\usepackage{xspace}

\ifx\colorrgbloaded\relax\let\next 
\else \let\colorrgbloaded\relax\let\next\relax \fi
\next
\catcode`\@=11\relax%
\newwrite\@unused%
\def\typeout#1{{\let\protect\string\immediate\write\@unused{#1}}}%
\def\beginrgbcolor#1#2#3{%
}
\def\endrgbcolor{%
}

\def\rgbcolor#1#2#3{\aftergroup\endrgbcolor%
	\beginrgbcolor{#1}{#2}{#3}}
\def\beginhsbcolor#1#2#3{%
    \special{ps::[begin]%
	 currenthsbcolor #1 255 div #2 255 div #3 255 div sethsbcolor}}
\def\endhsbcolor{%
	\special{ps::[end] sethsbcolor}}

\def\hsbcolor#1#2#3{\aftergroup\endhsbcolor%
	\beginhsbcolor{#1}{#2}{#3}}
\def\begincymkcolor#1#2#3#4{%
    \special{ps::[begin]%
	 currentcymkcolor #1 255 div #2 255 div %
		#3 255 div #4 255 div setcymkcolor}}
\def\endcymkcolor{%
	\special{ps::[end] setcymkcolor}}

\def\cymkcolor#1#2#3#4{\aftergroup\endcymkcolor%
	\begincymkcolor{#1}{#2}{#3}{4}}
\countdef\CCap=90
\countdef\Join=91
\def\InitCapJoinMatrix#1#2{%
\ifnum #1 >2 #2 > 2 #1 < 0 #2 < 0 {%
	\typeout{***bad argument range in function InitCapJoinMatrix}
}\else{%
	\CCap=#1\Join=#2%
}
\fi
}

\def\SetCapJoinMatrix#1#2{%
\ifnum #1 >2 #2 > 2 #1 < 0 #2 < 0 {%
	\typeout{***bad argument range in function InitCapJoinMatrix}
}\else%
	(\the\CCap ) cvi setlinecap (\the\Join ) cvi setlinejoin %
\fi
}
\def\fboxcolor#1#2#3#4{\leavevmode%
   \setbox\@tempboxa\hbox{#1}\@tempdima\fboxrule%
   \advance\@tempdima \fboxsep \advance\@tempdima\dp\@tempboxa%
   \beginrgbcolor{#2}{#3}{#4}%
   \hbox{\lower\@tempdima\hbox%
	{\vbox{\hrule\@height\fboxrule%
          \hbox{\vrule\@width\fboxrule\hskip\fboxsep%
	    \vbox{\vskip\fboxsep\endrgbcolor\box\@tempboxa%
		\beginrgbcolor{#2}{#3}{#4}\vskip\fboxsep}%
		   \hskip 1pt%
			 \hskip\fboxsep\vrule\@width\fboxrule}%
		   \hrule\@height\fboxrule}}}\endrgbcolor}%

\def\boxcolor#1#2#3#4{\leavevmode%
	\newbox\@myboxb\newdimen\@mydimb%
	\setbox\@myboxb\hbox{\hskip\fboxsep\hbox{#1}%
	\hskip\fboxsep}\@mydimb\fboxrule%
	\advance\@mydimb\dp\@myboxb%
	\beginrgbcolor{#2}{#3}{#4}%
	\hbox{
		\lower\@mydimb\hbox%
		{\vbox{\hsize=\ht\@myboxb\vsize=\wd\@myboxb%
		\hbox{\vrule\@width\wd\@myboxb%
		\hskip-\wd\@myboxb%
		\vbox{\endrgbcolor%
		\vskip\fboxsep\box\@myboxb\vskip\fboxsep}%
}}}}}%

\def\underlinecolor#1#2#3#4{%
	\newbox\@myboxbb\newbox\@myboxcc%
	\setbox\@myboxbb=\hbox{#1}%
	\setbox\@myboxcc=\hbox{\beginrgbcolor{#2}{#3}{#4}%
		   \vrule height .5pt depth 0pt width \wd\@myboxbb\endrgbcolor}
	\hbox{\raise -2.8pt\vbox{\offinterlineskip\@myboxbb\@myboxcc}}}

\countdef\rouge=112
\countdef\vert=113
\countdef\bleu=114
\countdef\increrouge=115
\countdef\increvert=116
\countdef\increbleu=117
\countdef\modulo=118
\countdef\sauverouge=119
\countdef\sauvevert=120
\countdef\sauvebleu=121
\countdef\maxrouge=122
\countdef\maxvert=123
\countdef\maxbleu=124
\def\initcolor#1#2#3#4#5#6#7{
	\rouge=#1\vert=#2\bleu=#3\modulo=#7%
	\sauverouge=#1\sauvevert=#2\sauvebleu=#3%
	\maxrouge=#4\maxvert=#5\maxbleu=#6}
\def\initinc#1#2#3{
	\increrouge=#1\increvert=#2\increbleu=#3}
\def\incrouge#1{%
	\global\advance\rouge by #1}
\def\incvert#1{%
	\global\advance\vert by #1}
\def\incbleu#1{%
	\global\advance\bleu by #1}
\def\decrouge#1{%
	\global\advance\rouge by -#1}
\def\decvert#1{%
	\global\advance\vert by -#1}
\def\decbleu#1{%
	\global\advance\bleu by -#1}
\def\incmodulo#1{
	\global\advance\modulo by #1}
\initcolor{20}{20}{20}{210}{210}{210}{1}%
\initinc{40}{40}{40}%

\def\echelons#1{\def\tete{}%
	\countdef\myinc=200%
	\myinc=1%
	\def\ECHELON##1{\ifx##1]%
	    \def\next{\tete}%
	    \else\ifnum\myinc>\modulo{%
			\ifnum\sauverouge<\maxrouge
				{\incrouge{\the\increrouge}}%
			\else
				{\decrouge{\the\increrouge}}%
			\fi
			\ifnum\sauvevert<\maxvert
				{\incvert{\the\increvert}}%
			\else
				{\decvert{\the\increvert}}%
			\fi
			\ifnum\sauvebleu<\maxbleu
				{\incbleu{\the\increbleu}}%
			\else
				{\decbleu{\the\increbleu}}%
			\fi
			\multiply\myinc by 0%
			\global\advance\myinc by 2%
		}\else{%
			\global\advance\myinc by 1%
		}\fi
		\ifnum\sauverouge<\maxrouge
			{\ifnum\rouge>\maxrouge{%
				\global\rouge=\sauverouge%
			}\fi}%
		\else
			{\ifnum\rouge<\maxrouge{%
				\global\rouge=\sauverouge%
			}\fi}%
		\fi%
		\ifnum\sauvevert<\maxvert
			{\ifnum\vert>\maxvert{%
				\global\vert=\sauvevert%
			}\fi}%
		\else
			{\ifnum\vert<\maxvert{%
				\global\vert=\sauvevert%
			}\fi}%
		\fi%
		\ifnum\sauvebleu<\maxbleu
			{\ifnum\bleu>\maxbleu{%
				\global\bleu=\sauvebleu%
			}\fi}%
		\else
			{\ifnum\bleu<\maxbleu{%
				\global\bleu=\sauvebleu%
			}\fi}%
		\fi%
		\edef\tete{\tete\beginrgbcolor{\number\rouge}{\number\vert}{\number\bleu}##1\endrgbcolor}%
		\let\next=\ECHELON\fi%
		\next}\ECHELON}

\countdef\sens=201%
\def\dentdescies#1{\def\tete{}%
	\countdef\myinc=200%
	\myinc=1%
	\def\DENTDESCIE##1{\ifx##1]%
	    \def\next{\tete}%
	    \else\ifnum\myinc>\modulo{%
			\ifnum\sens>0{%
				\incrouge{\the\increrouge}%
				\incvert{\the\increvert}%
				\incbleu{\the\increbleu}%
			}\else{%
				\decrouge{\the\increrouge}%
				\decvert{\the\increvert}%
				\decbleu{\the\increbleu}%
			}\fi%
			\global\myinc=2%
		}\else{%
			\global\advance\myinc by 1%
		}\fi
		\ifnum\rouge>\maxrouge{%
			\global\rouge=\maxrouge%
			\global\sens=0%
		}\fi%
		\ifnum\vert>\maxvert{%
			\global\sens=0%
			\global\vert=\maxvert%
		}\fi%
		\ifnum\bleu>\maxbleu{%
			\global\bleu=\maxbleu%
			\global\sens=0%
		}\fi%
		\ifnum\rouge<\sauverouge{%
			\global\rouge=\sauverouge%
			\global\sens=1%
		}\fi%
		\ifnum\vert<\sauvevert{%
			\global\sens=1%
			\global\vert=\sauvevert%
		}\fi%
		\ifnum\bleu<\sauvebleu{%
			\global\bleu=\sauvebleu%
			\global\sens=1%
		}\fi%
		\edef\tete{\tete\beginrgbcolor{\number\rouge}{\number\vert}{\number\bleu}##1\endrgbcolor}%
		\let\next=\DENTDESCIE\fi%
		\next}\DENTDESCIE}

\def\degrade#1{%
	\countdef\lg=220%
	\longueur[#1]%
	\ifnum\lg=0{%
		\typeout{macro degrade : parametre vide}}
	\else{%
	    \ifnum\lg>1{%
		\global\advance\lg by -1
		\ifnum\sauverouge<\maxrouge{%
			\increrouge=\maxrouge%
			\global\advance\increrouge by -\sauverouge}%
		\else{%
			\increrouge=\sauverouge%
			\global\advance\increrouge by -\maxrouge}%
		\fi
		\global\divide\increrouge by \lg%
		\ifnum\sauvevert<\maxvert{%
			\increvert=\maxvert%
			\global\advance\increvert by -\sauvevert}%
		\else{%
			\increvert=\sauvevert%
			\global\advance\increvert by -\maxvert}%
		\fi
		\global\divide\increvert by \lg%
		\ifnum\sauvebleu<\maxbleu{%
			\increbleu=\maxbleu%
			\global\advance\increbleu by -\sauvebleu}%
		\else{%
			\increbleu=\sauvebleu%
			\global\advance\increbleu by -\maxbleu}%
		\fi
		\global\divide\increbleu by \lg}%
	    \fi%
	    \echelons[#1]}%
	\fi}%

\def\longueur#1{%
	\lg=0%
	\def\LONGUEUR##1{%
		\ifx##1]%
			\def\next{}%
		\else%
			\global\advance\lg by 1%
			\let\next=\LONGUEUR%
		\fi
		\next}
	\LONGUEUR}

\def\PScouleur#1#2#3#4#5#6{%
\ifcase #6\or
\or
\or
\fi}

\def\CercleCouleurRGB#1#2#3#4#5#6{%
\dimen4=#4
\divide\dimen4 by 2
\hbox{\hsize\dimen4\hskip\dimen4
	\vbox to #4{\vskip\dimen4
		\PScouleur{#1}{#2}{#3}{#4}{#5}{#6}
	}%
      \hskip-\dimen4\hskip8pt
      \setbox1=\hbox{#1}%
      \dimen5=\wd1%
      \divide\dimen5 by 2%
      \raise\dimen5\vbox{\hbox{#1}}
}}

\def\CercleCouleur#1#2#3#4#5{%
\setbox3\hbox{#1}\newdimen\rayon
\ifdim\ht3>\wd3%
	\rayon=\ht3%
\else%
	\rayon=\wd3%
\fi%
\advance\rayon by 16pt
\ifnum #2 > 255 #3 <0
	\typeout{Les parametres de CercleCouleur sont incorrectes !!!}
\else
	\ifcase #4\or\CercleCouleurRGB{#1}{#2}{#3}{\the\rayon}{#5}{1}
			  \or\CercleCouleurRGB{#1}{#2}{#3}{\the\rayon}{#5}{2}
			  \or\CercleCouleurRGB{#1}{#2}{#3}{\the\rayon}{#5}{3}
	\fi
	\ifnum #4 > 3
	    \typeout{+++++Mauvais parametre dans Cercle Couleur : #4}
	\fi
\fi
}

\newdimen\xtrans\xtrans=0pt
\newdimen\ytrans\ytrans=0pt

\def\PSgenerale#1#2#3#4#5#6#7#8{%
}

\def\CercleGRGB#1#2#3#4#5#6#7#8#9{%
\dimen4=#9%
\divide\dimen4 by 2%
\dimen5\dimen4
\hbox to #9{\hskip\dimen5
	\vbox to #9{\vskip\dimen4%
		\PSgenerale{#2}{#3}{#4}{#5}{#6}{#7}{#8}{#9}%
	}%
	\hskip-\dimen5
	\setbox1=\hbox{#1}%
	\dimen5=#9\divide\dimen5 by 2%
	\dimen4=\ht1\divide\dimen4 by 2
	\advance\dimen5 by -\dimen4
	\raise\dimen5\vbox{\hbox{#1}}
}}

\def\CercleGeneral#1#2#3#4#5#6#7#8{%
\setbox3\hbox{#1}\newdimen\rayon%
\ifdim\ht3>\wd3%
	\rayon=\ht3%
\else%
	\rayon=\wd3%
\fi%
\advance\rayon by 16pt%
\ifnum #2 > 255 #5 <0 #3 > 255 #6 <0 #4 > 255 #7 <0%
	\typeout{Les parametres de CercleGeneral sont incorrects !!!}
\else
	\CercleGRGB{#1}{#2}{#3}{#4}{#5}{#6}{#7}{#8}{\the\rayon}%
\fi
}

\newif\ifhorizontal
\horizontaltrue  

\def\PSrectangle#1#2#3#4#5#6#7#8{%
\setbox6\hbox{#8}
}

\def\RectangleDegrade#1#2#3#4#5#6#7#8{%
\setbox3=\hbox{\hskip 8pt\vbox{\vskip 2pt\hbox{#1}\vskip 2pt}\hskip 8pt}
\ifnum #2 > 255 #5 <0 #3 > 255 #6 <0 #4 > 255 #7 <0
	\typeout{Les parametres de RectangleGeneral sont incorrects !!!}
\else
\hbox{\hskip-4pt
\hbox{\hsize\wd3%
	\vbox to \ht3{%
		\vskip-3pt%
		\PSrectangle{#2}{#3}{#4}{#5}{#6}{#7}{#8}{\copy3}
	}%
	\hskip 11pt%
	\raise 8pt\vbox{\hbox{#1}}}
}
\fi
}

\newif\ifdashps 
\dashpsfalse
\def\bezierps#1 \endbezierps#2#3#4#5#6#7{%
}

\def\ellipseps#1#2#3#4#5#6#7#8{%
}

\def\Line
#1 \endLine#2#3#4#5#6#7{%
}

\def\CercleHachure#1#2#3#4#5#6#7#8{%
\ifnum #4 > 180 #4 <0
    \typeout{---CercleHachure : le rayon #4 n'est pas <= 180 ni >= 0}
\else
\fi}

\def\CarreHachure#1#2#3#4#5#6#7#8{%
\ifnum #4 > 180 #4 <0
    \typeout{---CarreHachure : l'angle #4 n'est pas <= 180 ni >= 0}
\else
\fi}

\def\CourbeFermee#1 \endcourbe#2#3#4#5#6#7{%
\typeout{+++coordonnees: #1}
}
\def\filtre[ [ #1 #2 #3 ] ] {#1 #2 }
\newif\ifclosepath
\closepathtrue

\def\dc{\edef\t@vss{}\edef\b@vss{\vss}\edef\l@hss{}\edef\r@hss{\hss}}
\dc
\def\RotBis#1{\vbox to 0pt{\normalbaselines
	\t@vss
	\hbox to 0pt{\l@hss
		\hbox {#1}%
	\r@hss}%
	\b@vss}}
\def\fc{\edef\t@vss{\vss}\edef\b@vss{}\edef\l@hss{\hss}\edef\r@hss{}}
\def\mc{\edef\t@vss{\vss}\edef\b@vss{\vss}\edef\l@hss{\hss}\edef\r@hss{\hss}}

\def\ColleTexte#1#2#3#4#5#6#7#8#9{%
\typeout{***** limitation de ColleTexte : dvips 5.4 uniquement ****}
\countdef\lg=220%
\longueur[#1]\hss%
\courbebisbis[#1]}%
\def\courbebisbis#1{\def\tete{}%
	\def\COURBEBISBIS##1{\ifx##1]%
	    \def\next{\tete
}%
	    \else%
		\setbox20=\hbox{##1}%
		\edef\tete{\tete%
			\hss
\hss%
			\setbox21=\hbox{##1\hss}\box21%
			}\let\next=\COURBEBISBIS\fi%
			\next}\COURBEBISBIS}%

\countdef\BlancInterLettre=221%
\def\TraceTexte#1#2#3#4#5#6#7#8#9{%
\typeout{***** limitation de TraceTexte : dvips 5.4 uniquement ****}
\hss%
\special{ps: ) bidon %
	/D { 12.0 Resolution 300 div mul } def %
	/CY { Resolution 300 div } def %
	/MyCte { 0.35277 } def %
	/ancien-nt 0 def /ancien-y 0 def /ancien-x 0 def ( }\hss%
\special{ps: ) bidon %
	/fonctionX { n-t n-t n-t ax mul mul mul n-t n-t bx mul mul n-t cx mul %
		#2 add add add} def %
	/fonctionY { n-t n-t n-t ay mul mul mul n-t n-t by mul mul n-t cy mul %
		#3 add add add} def %
	/long #2 def %
	/blanc {\the\BlancInterLettre} def ( }\hss%
\hss%
\courbebis[#1]}%

\def\courbebis#1{\def\tete{}%
	\def\COURBEBIS##1{\ifx##1]%
	    \def\next{\tete\hss
}%
	    \else%
		\setbox20=\hbox{##1}%
		\edef\tete{\tete\hss%
\hss%
			\special{ps: ) bidon %
				MyFunction %
				{ fonctionX long le %
					{ /n-t n-t 0.001 add def } %
					{exit} ifelse } loop %
				/xx fonctionX def %
				/yy fonctionY def ( }\hss%
			\special{ps: ) bidon %
				currentpoint translate currentpoint %
				angle neg rotate neg exch neg exch %
				translate ( }\hss%
			\setbox21=\hbox{##1\hss}\box21%
			}\let\next=\COURBEBIS\fi%
			\next}\COURBEBIS}%

\newif\ifarrowheads
\newif\ifarrowheadleft
\newif\ifarrowheadright
\def\ArrowForm#1{
	\arrowheadsfalse
	\arrowheadleftfalse
	\arrowheadrightfalse
	\ifx#1h%
		\arrowheadstrue
	\else
		\ifx#1l%
			\arrowheadlefttrue
		\else
			\ifx#1r%
				\arrowheadrighttrue
			\else
				\typeout{*** bad parameter in function ArrowForm}
			\fi
		\fi
	\fi
}

\countdef\Psymbol=205\countdef\Ssymbol=206

\def\PositionPsymbol#1{
	\ifnum#1>100 #1<0
		\typeout{*** bad parameter in function PositionPsymbol}
	\else
		\Psymbol=#1
	\fi
}

\def\PositionSsymbol#1{
	\ifnum#1>100 #1<0
		\typeout{*** bad parameter in function PositionSsymbol}
	\else
		\Ssymbol=#1
	\fi
}

\newif\ifinnerbox\innerboxtrue
\def\TroisDbox#1#2#3#4#5#6#7#8{%
\ifnum#1=0
\typeout{===== Le 2eme Parametre de \TroisDBox doit etre <>0 =====}
\else
\fi}

\def\plotfunction#1#2#3#4#5#6#7{%
}

\newcount\pasx 
\newcount\pasy 
\newcount\mycinterone\newcount\mycintertwo
\newcount\vinitx\newcount\vincx
\newcount\vinity\newcount\vincy
\def\GridNumPuiss(#1,#2)(#3,#4)(#5,#6)#7#8#9{%
	\put(#1,#2){\vector(1,0){#3}}
	\put(#1,#2){\vector(0,1){#4}}
	\mycinterone=#2
	\advance\mycinterone by -5
	\mycintertwo=#3
	\divide\mycintertwo by \pasx\advance\mycintertwo by 1
	\multiputnum(#1,\number\mycinterone)(\number\pasx,0)%
		{\number\mycintertwo}{\number\vinitx}{\number\vincx}
	\mycintertwo=#4
	\divide\mycintertwo by \pasy\advance\mycintertwo by 1
	\multiputpuissance(\number\mycinterone,#2)(0,\number\pasy)%
		{\number\mycintertwo}{\number\vinity}{\number\vincy}
	\mycinterone=#1
	\advance\mycinterone by -2
	\multiput(\number\mycinterone,\number\pasy)(0,\number\pasy)%
		{\number\mycintertwo}{\traith{3}}
	\mycinterone=#3
	\divide\mycinterone by \pasx\advance\mycinterone by 1
	\mycintertwo=#2
	\advance\mycintertwo by -2
	\multiput(\number\pasx,\number\mycintertwo)(\number\pasx,0)%
		{\number\mycinterone}{\traitv{4}}
	\put(#1,#2){\Grille(#3,#4)(#5,#6){#7}{#8}{#9}}
}

\def\GridNumNum(#1,#2)(#3,#4)(#5,#6)#7#8#9{%
	\put(#1,#2){\vector(1,0){#3}}
	\put(#1,#2){\vector(0,1){#4}}
	\mycinterone=#2
	\advance\mycinterone by -5
	\mycintertwo=#3
	\divide\mycintertwo by \pasx\advance\mycintertwo by 1
	\multiputnum(#1,\number\mycinterone)(\number\pasx,0)%
		{\number\mycintertwo}{\number\vinitx}{\number\vincx}
	\mycintertwo=#4
	\divide\mycintertwo by \pasy\advance\mycintertwo by 1
	\multiputnum(\number\mycinterone,#2)(0,\number\pasy)%
		{\number\mycintertwo}{\number\vinity}{\number\vincy}
	\mycinterone=#1
	\advance\mycinterone by -2
	\multiput(\number\mycinterone,\number\pasy)(0,\number\pasy)%
		{\number\mycintertwo}{\traith{3}}
	\mycinterone=#3
	\divide\mycinterone by \pasx\advance\mycinterone by 1
	\mycintertwo=#2
	\advance\mycintertwo by -2
	\multiput(\number\pasx,\number\mycintertwo)(\number\pasx,0)%
		{\number\mycinterone}{\traitv{4}}
	\put(#1,#2){\Grille(#3,#4)(#5,#6){#7}{#8}{#9}}
}

\def\Grille(#1,#2)(#3,#4)#5#6#7{%
}

\def\Fleche#1 \endfleche#2#3#4#5#6#7#8#9{%
\arrowheadsfalse%
\arrowheadleftfalse%
\arrowheadrightfalse%
}
		
\def\Cotes(#1,#2)(#3,#4)[#5,#6][#7,#8]|#9|{%
}

\def\EtatTransition#1{
}

\newif\ifarrowheads\arrowheadsfalse
\newif\ifarrowheadleft\arrowheadleftfalse
\newif\ifarrowheadright\arrowheadrightfalse
\newif\ifencore\encoretrue
\def\ArrowForm#1{%
	\arrowheadsfalse%
	\arrowheadleftfalse%
	\arrowheadrightfalse%
	\def\bidonarrow{#1 }%
	\def\hh{h }\def\rr{r }\def\ll{l }%
	\loop\message{Initialisation de la forme des fleches}%
	\ifx\bidonarrow\hh%
		\arrowheadstrue\encorefalse%
	\else%
		\ifx\bidonarrow\ll%
			\arrowheadlefttrue\encorefalse%
		\else%
			\ifx\bidonarrow\rr%
				\arrowheadrighttrue\encorefalse%
			\else%
				\message{*** bad parameter in function ArrowForm (h l r)}%
				\read16 to \bidonarrow%
				\encoretrue%
			\fi%
		\fi%
	\fi%
	\ifencore%
	\repeat%
}

\def\FlecheSpeciale#1 \endflechespeciale#2#3#4#5#6#7#8#9{%
\ArrowForm{#2}\Fleche #1 \endfleche{100}{#3}{#4}{#5}{#6}{#7}{#8}{#9}%
}

\def\TraceArrow(#1,#2)(#3,#4)|#5|{%
}%

\newif\ifexterne\externetrue
\def\CotesFleche(#1,#2)(#3,#4)[#5,#6][#7,#8]|#9|{%
\Cotes(#1,#2)(#3,#4)[#5,#6][#7,#8]|#9|\TraceArrow(#1,#2)(#3,#4)|#9|%
}

\newcount\multicnt 
\newdimen\xdim
\newdimen\ydim
\newcount\fechelle

\def\whilenoop#1{}
\def\whilenum#1\do #2{\ifnum #1\relax #2\relax\iwhilenum{#1\relax 
     #2\relax}\fi}
\def\iwhilenum#1{\ifnum #1\let\nextwhile=\iwhilenum 
         \else\let\nextwhile=\whilenoop\fi\nextwhile{#1}}

\def\whiledim#1\do #2{\ifdim #1\relax#2\iwhiledim{#1\relax#2}\fi}
\def\iwhiledim#1{\ifdim #1\let\nextwhile=\iwhiledim 
        \else\let\nextwhile=\whilenoop\fi\nextwhile{#1}}

\long\def\multiputnum(#1,#2)(#3,#4)#5#6#7{\killglue\multicnt=#5\relax
\xdim=#1\unitlength%
\ydim=#2\unitlength%
\fechelle=#6
\whilenum \multicnt > 0\do%
{\raise\ydim\hbox to 0mm{\setbox2=\hbox{{\the\fechelle}}%
\dimen1=\wd2\divide\dimen1 by 2\wd2=\dimen1%
\dimen1=\ht2\divide\dimen1 by 2\ht2=\dimen1%
\kern-\wd2\kern\xdim\raise-\ht2\hbox{\the\fechelle}\hss}\advance\multicnt-1%
\advance\fechelle#7%
\advance\xdim#3\unitlength\advance\ydim #4\unitlength}\ignorespaces}

\long\def\multiputpuissance(#1,#2)(#3,#4)#5#6#7{\killglue\multicnt=#5\relax
\xdim=#1\unitlength
\ydim=#2\unitlength
\fechelle=#7
\whilenum \multicnt > 0\do%
{\raise\ydim\hbox to 0mm{\setbox2=\hbox{$#6^{\the\fechelle}$}%
\dimen1=\wd2\divide\dimen1 by 2\wd2=\dimen1%
\dimen1=\ht2\divide\dimen1 by 2\ht2=\dimen1%
\kern-\wd2\kern\xdim\raise-\ht2\hbox{$#6^{\the\fechelle}$}\hss}%
\advance\multicnt-1%
\advance\fechelle#7%
\advance\xdim#3\unitlength\advance\ydim#4\unitlength}\ignorespaces}

\def\killglue{\unskip\whiledim \lastskip >0mm\do{\unskip}}

\def\traith#1{\hbox{\vrule height .4pt depth 0pt width #1mm}}
\def\traitv#1{\hbox{\vrule height #1mm depth 0pt width 0.4pt}}

\def\LogoFlecheOne#1#2#3{%
}

\newdimen\dimeminter
\def\LeftLogoFlecheOne{\leavevmode\vadjust{\setbox1=\vtop{\hsize 10mm %
	\parindent=0pt\baselineskip=9pt\rightskip=4mm plus %
	4mm\LogoFlecheOne{0.04}{0.9}{0.9}}%
	\dimeminter=\textwidth \advance\dimeminter by -\leftmargin
	\hbox to \dimeminter{\rlap{%
		\kern\leftmargin\kern-15mm\smash{\raise6pt\box1}}\hfill}}}

\def\LogoFlecheTwo#1#2#3{%
}

\def\LeftLogoFlecheTwo{\leavevmode\vadjust{\setbox1=\vtop{\hsize 10mm %
	\parindent=0pt\baselineskip=9pt\rightskip=4mm plus %
	4mm\LogoFlecheTwo{0.02}{0.55}{0.55}}%
	\dimeminter=\textwidth \advance\dimeminter by -\leftmargin
	\hbox to \dimeminter{\rlap{%
		\kern\leftmargin\kern-15mm\smash{\raise6pt\box1}}\hfill}}}

\def\SunStation#1#2#3{
}

\def\SunDisk#1#2#3{
}

\def\SunServer#1#2#3{
}

\def\SunRouter#1#2#3{
}
\catcode`\@=12\relax%

\newcommand{\Mbh}{\ensuremath{M_{\bullet}}}

\newcommand{\ML}[1][]{\ensuremath{\Upsilon_{#1}}}

\newcommand{\Msun}{\ensuremath{M_{\sun}}}

\newcommand{\eg}{\rm {e.g.,~}}
\newcommand{\ie}{\rm {i.e.,~}}
\newcommand{\etal}{\rm {et al.~}}

\newcommand{\sauron}{\texttt{SAURON}\xspace}
\newcommand{\oasis}{\texttt{OASIS}\xspace}

\newcommand{\NE}{\ensuremath{{\rm N}_E}}
\newcommand{\NLz}{\ensuremath{{\rm N}_{L_z}}}
\newcommand{\Ncomp}{\ensuremath{{\rm N_{comp}}}}

\begin{document}

\title[On the reliability of the black hole mass and mass to light ratio determinations]{On
the reliability of the black hole mass and mass to light ratio 
determinations with Schwarzschild models}

\author[N. Cretton and E. Emsellem]{Nicolas Cretton$^{1}$\thanks{E--mail:
ncretton@eso.org; emsellem@obs.univ-lyon1.fr} and 
Eric Emsellem$^{2}$\footnotemark[1]\\
$^{1}$ESO, Karl Schwarzschild Strasse 2, 85748 Garching bei M\"unchen, Germany\\
$^{2}$CRAL--Observatoire, 9 Avenue Charles--Andr\'e, 69230 Saint--Genis--Laval, France}

\date{Accepted .... Received ...}

\pagerange{\pageref{firstpage}--\pageref{lastpage}} \pubyear{2003}

\maketitle

\label{firstpage}

\begin{abstract}
In this Letter, we investigate the claim of Valluri \etal (2003),
namely that the use of Schwarzschild dynamical (orbit--based) models
leads to an indeterminacy regarding the estimation of the free
parameters like the central black hole mass and the stellar
mass--to--light ratio of the galaxy under study. We examine this issue
with semi--analytic two--integral models, which are not affected by
the intrinsic degeneracy of three--integral systems.  We however
confirm the Valluri et al. result, and observe the so--called widening
of the $\chi^2$ contours as the orbit library is expanded.  We also
show that, although two--dimensional data coverage help in
constraining the orbital structure of the modelled galaxy, it does not
in principle solve the indeterminacy issue, which mostly originates
from the discretization of such an inverse problem. We show that
adding regularization constraints stabilizes the confidence level
contours on which the estimation of black hole mass and stellar
mass--to--light ratio are based.  We therefore propose to
systematically use regularization as a tool to prevent the solution to
depend on the orbit library.  Regularization, however, introduces an
unavoidable bias on the derived solutions.  We hope that the present
Letter will trigger some more research directed at a better
understanding of the issues addressed here.
\end{abstract}


\section{Introduction}

A recent paper by Valluri, Merritt \& Emsellem (2003; hereafter VME03)
demonstrated that axisymmetric orbit--based modelling algorithms have
only a limited ability to recover two fundamental parameters
characterizing the total mass distribution, namely the stellar
mass--to--light ratio $\ML{}$ and the black hole (BH) mass
$\Mbh$. These authors showed that if the number of orbits is
increased, the confidence contours defining the best fit values for
[\Mbh; $\ML{}$] and their associated error bars widen up, hence
significantly increasing the uncertainty of the BH measurement (see
\eg their Figure~3). In the case of M32, they showed that a range of
models with $\Mbh$ values between $1 \times 10^6 \Msun$ and $6 \times
10^6 \Msun$ provide equally good fit to the data discussed by van der
Marel et al. (1998).  This contrasts with the claim of other groups,
\eg ``the range of black hole mass uncertainties is from 5\% to 70\%,
with an average uncertainty around 20 \%" (Gebhardt 2003).

We have investigated this question using compo\-nent--based $f(E,L_z)$
models, where $f(E,L_z)$ is the distribution function (DF) depending
only on the energy $E$ and the vertical angular momentum $L_z$
(Sect.~\ref{model_and_data}). We confirm the same widening effect
reported by VME03, albeit with two--integral models instead of
three--integral models. In Sect.~\ref{components} we argue that the
specific effect we describe here is due to the discretization of
phase--space (\ie the sampling into a finite number of orbits or
components) and is therefore not related to the peculiar form of the
distribution function.

We illustrate our claim within a realistic context, namely using the
luminous mass model, as well as the \sauron and
\oasis\footnote{\sauron and \oasis are integral--field spectrographs}
setups (Bacon et al. 2001, de Zeeuw et al. 2002) for NGC~3377 (see
Copin \etal 2003, hereafter C03): these are described in
Sect.~\ref{model_and_data}.

We have considered two options in order to examine the reliability of
the $\ML{}$ and $\Mbh$ determinations. First, we have investigated
whether the spatial extension of the data helps to better constrain
the model (Sect.~\ref{2D_extension}).  Second, we have included
regularization in the fitting process (Sect.~\ref{Regularization}) and
show that it provides a way to stabilize the widening effect in the
two--integral case.
 
\begin{figure*}
\resizebox{1.0\textwidth}{!}{\includegraphics{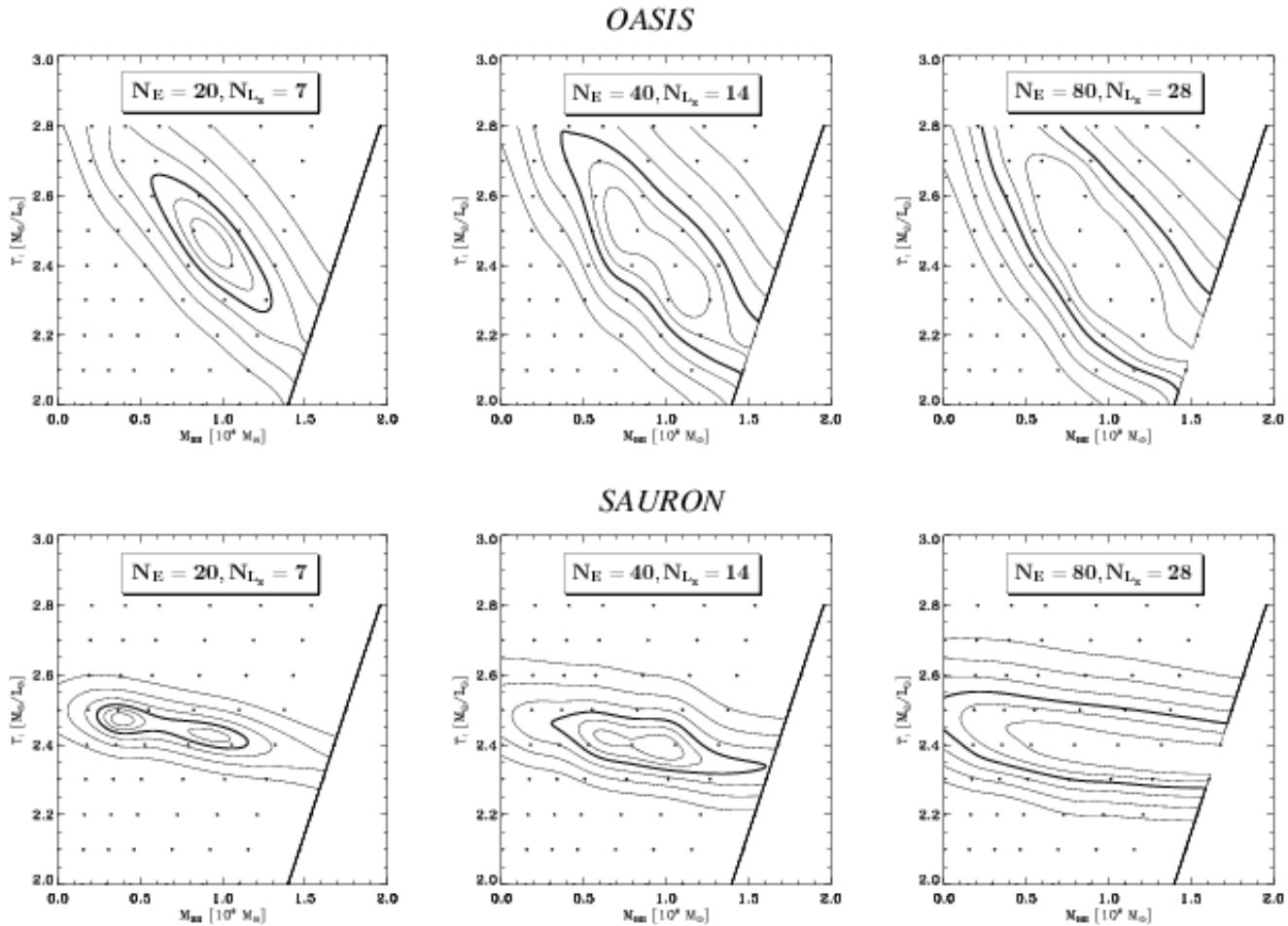}}
\caption{$\chi^2$--contours without any regularization. Models
including only the \oasis constraints are shown on the top row,
whereas the bottom row displays models with only the \sauron
constraints.  The number of components increases from left to right
(see insets). The third, thicker, contour (counting from the innermost
one) corresponds to the formal 99.7\% confidence level with two
degrees of freedom.}
\label{contours_non_regul}
\end{figure*}

\begin{figure*}
\resizebox{1.0\textwidth}{!}{\includegraphics{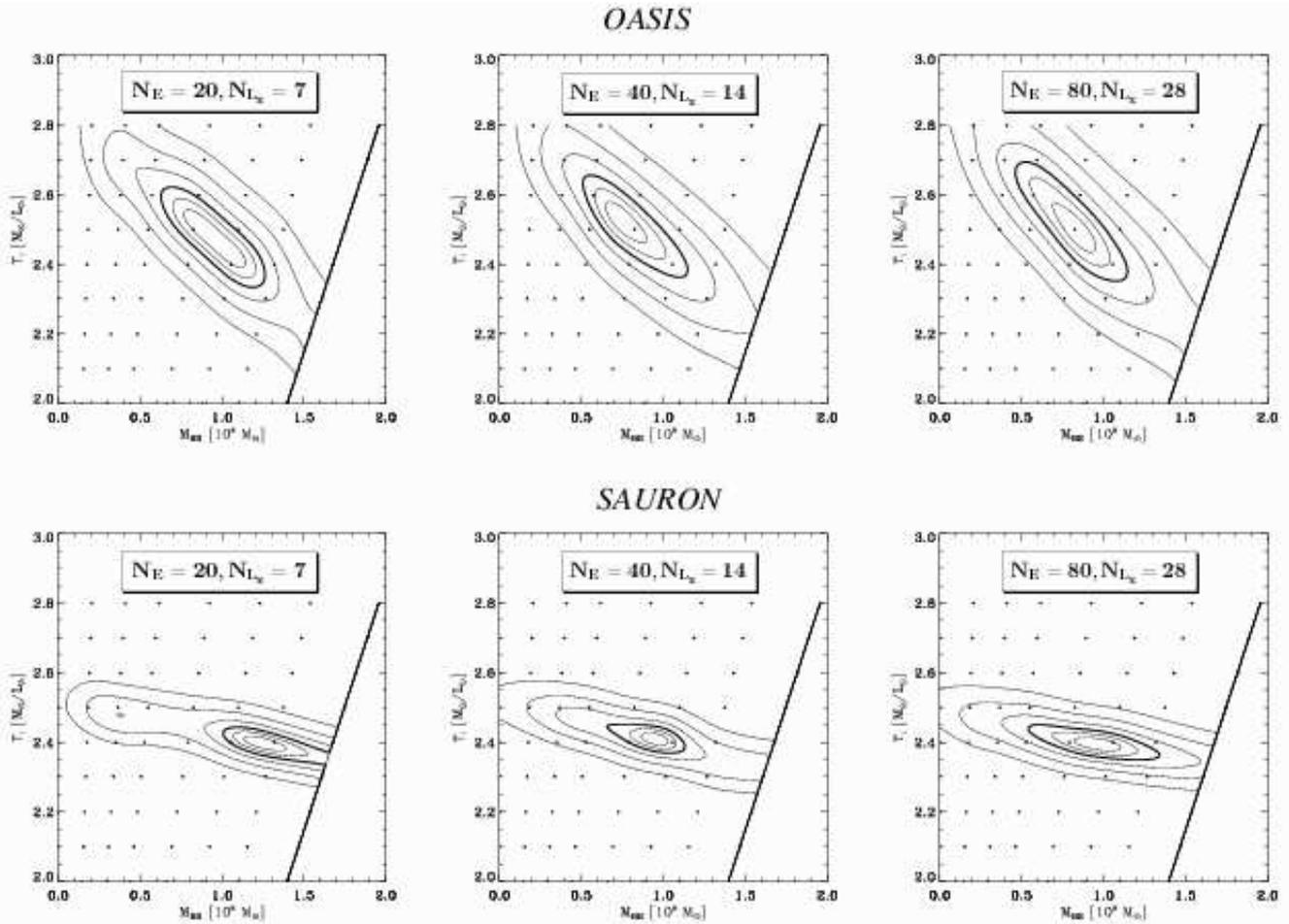}}
\caption{Same as figure~\ref{contours_non_regul} but for models including 
regularization}
\label{contours_regul}
\end{figure*}

\section{Input constraints and dynamical modelling}
\label{model_and_data}

\subsection{A two--integral model}

One of the advantages of working with a two--integral DF is
uniqueness: there is a unique $f(E,L_z)$ DF that matches
simultaneously the mass density, which is related to the even part of
the DF, and the mean velocity field which is fully specified by the
odd part of the DF (see \eg Merritt 1996b). There is therefore no
intrinsic degeneracy in recovering the input parameters [\Mbh;
$\ML{}$] for the two--integral models. Note, however, that in the
three--integral case the model has some additional freedom to adjust
its orbital anisotropy (e.g., in the three--integral case, the radial
velocity dispersion can be different than the vertical one). Thus,
three--integral models with [\Mbh; $\ML{}$] values very different from
the true input values may still be found to fit the data equally well
(see VME03). Because of this extra freedom, the confidence level
contours will always be larger in the three--integral case.

Another advantage of the two--integral models is that we can easily
check directly their entire dynamical structure by comparing the
reconstructed DF with the true input DF. In our view, these two
properties make the two--integral case a genuine {\em test case}.

\subsection{The mass model and kinematic constraints}
\label{constraints}

For the luminous density distribution we use our Multi--Gaussian
Expansion (MGE) model of NGC~3377 (see C03, and Emsellem \etal 1994).
The stellar $\ML{_I}$ has been fixed to 2.4 and a dark component in
the form of a $10^8$~\Msun~black hole has been included at the center:
this fixes the gravitational potential of our axisymmetric model.

Except for a few notable cases (see e.g., Cappellari \etal 2002;
Verolme \etal 2002, C03), the vast majority of BH determinations are
based on one--dimensional long--slit spectra, leaving most of the
$(x',y')$ sky plane unconstrained. One might reasonably expect that
data densely covering the projected plane of the sky $(x',y')$ would
help to constrain the models. This is indeed emphasized in Verolme
\etal (2002) and in C03, where it is shown that the estimates of the
free parameters of the models (e.g., inclination, $\Mbh$, $\ML{}$) are
significantly better constrained when two--dimensional photometric and
kinematic data are included. We have therefore tested two different
regimes, namely with two--dimensional constraints having different
spatial extensions (see C03; Bacon et al. 2001):
\begin{enumerate}
\item a central dataset, with the characteristics of the \oasis setup 
covering the inner $4''$ by $2''$ and,
\item a spatially extended dataset, with the full \sauron field
of view covering $46''$ by $42''$.
\end{enumerate}

From the MGE mass model, we have generated artificial kinematic
constraints assuming a semi-isotropic distribution function $f(E,L_z)$
using the Hunter--Qian (1993) method (HQ hereafter). The even part of
such a distribution function (DF) is uniquely specified by the MGE
mass model, and we chose the odd part so as to roughly mimic the
observed \sauron and \oasis kinematics of NGC~3377 (C03).  The
observables of the HQ model (\ie the line--of--sight velocity
profiles, hereafter VPs) are then computed for both setups: they are
convolved with the appropriate point spread functions and binned into
the respective spatial elements, and will be used to constrain the
dynamical models\footnote{In the remainder of this paper, ``\sauron
constraints'' refer to the artificial data derived from the HQ method,
using the \sauron setup as in C03 (pixel size, Point Spread Function,
etc).}.

\subsection{The $f(E,L_z)$ component--based Schwarzschild models}
\label{components}

We now turn to the building of a Schwarzschild dynamical model
constrained by the fake photometric and kinematic datasets described
in Sect.~\ref{constraints}.  We use the $f(E,L_z)$ component--based
models of Cretton \etal 1999 (hereafter C99).  These authors described
how to derive two--integral models (with or without central black
holes) using only these $f(E,L_z)$ components. These models
were fully specified by reconstruction of the
corresponding DFs (see their Figs. 6, 7 and 8), which were compared to
the true underlying DFs obtained via the HQ method (see Verolme \& de
Zeeuw 2002 for similar computations).
Following C99, and using the gravitational potential
corresponding to the MGE mass model (Sect.~\ref{model_and_data}), we
have created a library of two--integral components by sampling the
energy $E$ with \NE~values and, at each energy, by sampling the
angular momentum $L_z$ with \NLz values. The final solutions are
obtained with a non--negative least square algorithm (NNLS, see C99
and Lawson \& Hanson 1974 for details).

In the next Sections, we show that the widening effect reported
by VME03 is also present with two--integral models, despite the absence
of an intrinsic degeneracy for such models.

\section{Spatial extension}
\label{2D_extension}

Figure~\ref{contours_non_regul} shows the $\chi^2$ contours obtained
without regularization using 1. only the \oasis constraints, covering
the central parts (top row), or 2.  only the \sauron constraints with
a large field of view (bottom row).

The \oasis contours exhibit a clear correlation between the $\Mbh$ and
$\ML{_I}$: a higher black hole mass needs to be compensated by a lower
$\ML{_I}$.  With their spatial extension, the \sauron--like
observables provide a much better constraint on $\ML{_I}$, but
obviously fail to restrict the allowed range for $\Mbh$.  This is an
expected behaviour for such models, as discussed in details in C03.
What is more surprising is that for both cases, when the number of
components \Ncomp\ is increased, a larger region of the $[\Mbh;
\ML{_I}]$ plane is enclosed by the 99.7\% confidence level ($3\sigma$
error in two degrees of freedom, which corresponds to $\Delta\chi^2 =
11.8$, see Press \etal 1992). We have increased \Ncomp\ by up to a
factor of 16 and still, we see no sign of convergence. Using our
two--integral models, we therefore see an effect similar to the one
reported by VME03.

The wider field of view does not seem to stop the $\chi^2$--contours
from getting broader; as emphasized by the bottom row of
Fig.~\ref{contours_non_regul}, it merely slows the effect down.
However for the \sauron setup, we can still observe a significant
increase in the allowed range of $\ML{_I}$ (\ie within the $3\sigma$
contour), particularly when $\NE = 80$ and $\NLz = 28$.

\begin{figure}
\resizebox{0.5\textwidth}{!}{\includegraphics{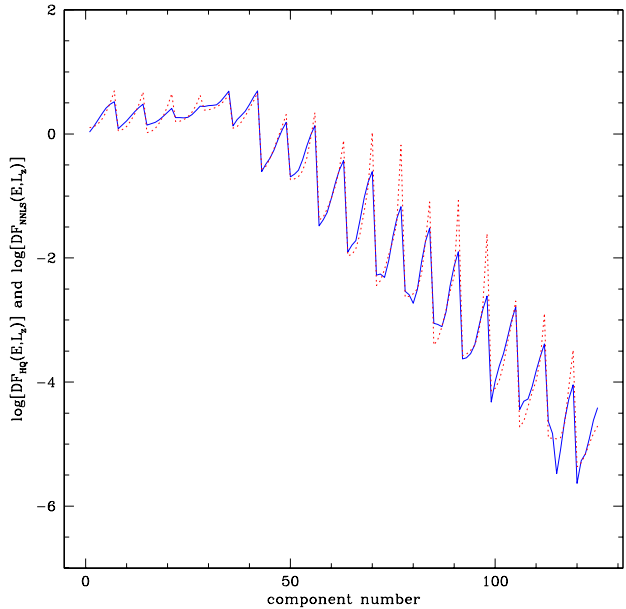}}
\caption{Distributions functions: the dotted line is the ``true''
${\rm DF_{HQ}}(E,L_z)$ derived using the HQ algorithm and the full
line is the reconstructed ${\rm DF_{NNLS}}(E,L_z)$ of the model with
$[\Mbh; \ML{_I}]$=$[10^8; 2.4]$. }
\label{DF_plot}
\end{figure}

\section{Towards Regularization}
\label{Regularization}

In the absence of regularization, the weights of many two--integral
components are set to zero by the NNLS algorithm: only ${\rm N}_{\rm
const}$ components can have a non zero weight, where ${\rm N}_{\rm
const}$ is the number of constraints (see C99). Such
a behaviour has been emphasized by e.g., Cappellari \etal (2002),
and Verolme \& de Zeeuw (2002). 

By increasing \Ncomp\ one improves the sampling of the integral 
space, which also corresponds to a better spatial sampling. In this
sense, with a bigger library NNLS has more flexibility to fit the
(same) data. An equally good fit to the data can then be obtained
using a different model than the true input one. A larger fraction of
the models in [\Mbh; $\ML{}$]--space are now acceptable, because NNLS
has more freedom to select the appropriate orbits. The discretization
of the phase space implies that the problem of recovering the orbital
weights become ill--conditioned. We believe the same effect occurs
with three--integral orbit--based models and suggest that it is partly
at the root of the contour widening problem (VME03).

Including regularization forces all the components to have a non zero
weight and therefore greatly reduces (or eliminates) this freedom.
Furthermore, with such jagged weight distributions, the solutions
without regularization are definitely far away from physical
solutions. We do not claim that our way of performing regularization
is optimal or represent in any sense what is taking place in real
galaxies. As emphasized in Sect.~\ref{2D_extension} and
in VEM03, unregularized solutions will depend on the size of the 
orbit library used in the analysis.  However, regularization introduces 
an unavoidable bias in the solutions the algorithm will find.

Regularization is done here in the same way as described in C99, \ie
by minimizing the second derivatives of the weights on a grid in
Integral space.

\subsection{Results}
\label{results} 

Figures~\ref{contours_regul} shows the results (as $\chi^2$ contours)
of using the \oasis (top row) or \sauron (bottom row) constraints,
when regularization is included.  The amount of regularization was set
such as to optimally reproduce the input HQ--DF (see Fig.~8 of
C99). Some other schemes to tune the optimal regularization parameter
exist (\eg generalized cross--validation, Wahba 1990; Merritt 1997).

In both \sauron and \oasis setups, the contours are stable in the
sense that, while increasing \Ncomp\ by up to a factor 16, they do not
show the widening effect observed without regularization. This can be
intuitively understood by examining the size of the matrix problem
which NNLS is trying to solve. Without regularization, the input
matrix containing the observables for each component has \Ncomp\
columns (number of components) and ${\rm N}_{\rm const}$ lines (number
of constraints). When \Ncomp\ is increased reaching values
significantly larger than ${\rm N}_{\rm const}$, the problem becomes
strongly underconstrained, and most of the resulting weights are set
to zero (see above).  When regularization is introduced, the number of
columns stays constant, while the number of lines becomes $\Ncomp +
{\rm N}_{\rm const} \times {\rm N}_{\rm Int}$ ( ${\rm N}_{\rm Int}$ is
the number of integral of motions, equal to 2 in the present
two--integral case). As \Ncomp\ is increased, the numbers of columns
and lines of the matrix increase simultaneously, their ratio
converging to $ 1/ {\rm N}_{\rm Int}$. Although it depends on the
regularization scheme used, these additional constraints can thus
qualitatively change the nature of the inverse problem.

We should emphasize here that the results presented in
this Section are no proof that the contours will not get wider for 
a much larger \Ncomp. Nevertheless, it is clearly an improvement 
over the unregularized case, the results of which depend on
the size of the orbit library. 

\subsection{Distribution functions}
\label{df}

A more conclusive test is to examine the DF corresponding to the
weights recovered by the least--square routine.  Figure~\ref{DF_plot}
compares the reconstructed ${\rm DF_{NNLS}}$ obtained from the NNLS
fit to the input ${\rm DF_{HQ}}$ derived from HQ. For each value of
the Energy, $L_z$ runs monotonically from $L_{z,{\rm min}}$ to
$L_{z,{\rm max}}$, covering ${\rm N}_{L_z}$ (=7 here) components. The
next component corresponds to the next value of the energy,
etc. Overall, the agreement is good, except for discrepancies at the
peaks near $L_{z,{\rm max}}$. A similar test presented by C99 on a
fake model of M~32 showed a significantly better fit (with RMS
residuals of the order of a few percents). The difference lies
certainly in the contribution of a disk--like component in the fake
model of NGC~3377, which would probably require a better sampling in
${L_z}$.  It is likely that some more advanced regularization scheme
could improve the recovery of these steep gradients in the DF, but
this is outside the scope of the present paper.

\section{Comparison with other models}

Gebhardt (2003) claimed that increasing the number of orbits does not
produce the contour widening effect described in VME03 and in the
present paper. The cause for this discrepancy is, at the moment,
unclear to us.

Models with and without regularization look surprisingly si--milar
when \NE=20 and \NLz=7, which are typical values used in many
orbit--based dynamical models (see \eg vdM98, C99, Cretton \& van den
Bosch 1999, Verolme \etal 2002, Cappellari \etal 2002).  We suggest
that this coincidence has (at least) two consequences. Firstly, this
could trigger the misconception that regularization does not affect
the indeterminacy in the estimation of [\Mbh; $\ML{}$], since the
$\chi^2$ contours (with or without regularization) are similar.
Secondly, this could explain the good apparent agreement sometimes
found between regularized and unregularized codes. If this comparison
had been performed using a much larger number of orbits (in the
unregularized case), we predict that a different conclusion would have
been reached.

\section{conclusion}

In this Letter, we have used two--integral models to test the claim
made by VME03 that $\chi^2$ contours derived using the Schwarzschild
orbit-based technique widen as the orbit library is expanded.
\begin{itemize}
\item We confirm this effect with two--integral
models for which the intrinsic degeneracy discussed in VME03 does not exist.
\item We show that the widening of the $\chi^2$ contours occurs even
when we use two--dimensional kinematics constraints.
\item We suggest that the origin of such an effect partly lies in the
discretization of phase space.
\item We finally propose to systematically use regularization as a way
to stabilize the $\chi^2$ diagrams produced by such techniques, so that
the obtained solution does not significantly depend on the input orbit library.
\end{itemize}

Regularization schemes have been often advocated to derive stable
solutions of ill--conditioned inverse problems (\eg Golub \etal 1999,
and references therein; Merritt 1996b).  The main worry associated with
regularization is that it biases the solution found by the fitting
algorithm (see \eg Merritt, 1996a and Dehnen, 2001 for similar
questions in N--body codes). This bias may indeed not correspond to
the true characteristics of the object under study.  The input
two--integral DF we chose for the present test is, by construction, a
smooth (semi--analytical) function of the energy and angular
momentum. Our analysis shows that in this case a simple regularization
scheme with the right choice of smoothing parameter can be applied to
recover the full (two--integral) DF reasonably well (see also C99 for
a similar demonstration on a less complex DF). When modelling real
galaxies, however, there is no obvious way to select the proper level
of smoothing, an incorrect choice leading to either degenerate
solutions (if undersmoothed; see Fig.~\ref{contours_non_regul}) or
biased solutions (if oversmoothed). The sampling of the orbit library
may be further optimized such as to ensure a relatively homogeneous
coverage of the presumed components of the galaxy (core, disk, bulge,
etc). As mentioned in Sect.~\ref{df}, it is likely that an adaptive
regularization scheme can be designed, \eg taking into account some a
priori knowledge (or preconception) we have on a galaxy (\eg the
presence of a bright disk).

The stability of regularized solutions however comes at the cost of an
unavoidable bias. The resulting best fit parameters and confidence
range for [\Mbh; $\ML{}$] in a galaxy may not necessarily reflect
their true values.  We emphasize that, in this Letter, we only treat
the indeterminacy caused by the discrete nature of the inverse problem
and not the additional degeneracy associated with 3--integral models,
as reported by VME03. We therefore hope that the present Letter will
trigger further studies to address these issues.

\subsection*{Acknowledgments}

We are grateful to Karl Gebhardt, Hans--Walter Rix, Michele Cappellari, 
Monica Valluri and David Merritt for many fruitful discussions.

\vspace{-0.5cm}




\clearpage


\begin{thebibliography}{}
\bibitem[Bacon et al.(2001)]{2001MNRAS.326...23B} Bacon, R.~et al.\ 2001, 
MNRAS, 326, 23 
\bibitem[Cappellari et al.(2002)]{2002ApJ...578..787C} Cappellari, M., 
Verolme, E.~K., van der Marel, R.~P., Kleijn, G.~A.~V., Illingworth, G.~D., 
Franx, M., Carollo, C.~M., \& de Zeeuw, P.~T.\ 2002, ApJ, 578, 787 
\bibitem[Copin, Cretton \& Emsellem 2003]{C03}
  Copin, Y., Cretton, N. \& Emsellem, E. 2003,
  A\&A, in press (astro-ph/0311387, C03)
\bibitem[Cretton et al. 1999]{Cretton99}
  Cretton, N., de Zeeuw, P. T., van der Marel, R. P., \&
  Rix, H.--W. 1999, 
  ApJS, 124, 383
\bibitem[Cretton \& van den Bosch 1999]{CrettonB99}
  Cretton, N. \& van den Bosch, F. C. 1999,
  ApJ, 514, 704
\bibitem[Dehnen 2001]{2001MNRAS.324..273D} Dehnen, W., 2001, MNRAS, 324, 273
\bibitem[de Zeeuw et al.(2002)]{2002MNRAS.329..513D} de Zeeuw, P.~T.~et 
  al.\ 2002, MNRAS, 329, 513 
\bibitem[Emsellem, Monnet \& Bacon 1994]{EMB94}
  Emsellem, E., Monnet, G., \& Bacon, R. 1994, 
  A\&A, 285, 723
\bibitem[Gebhardt et al.(2003)]{2003ApJ...583...92G} Gebhardt, K.~et al.\ 
  2003, ApJ, 583, 92 
\bibitem[Gebhardt (2003)]{} Gebhardt, K.\, 2003, Carnegie
  Observatories Astrophysics Series, Vol. 1: Coevolution
  of Black Holes and Galaxies, ed. L. C. Ho (Cambridge:
  Cambridge Univ. Press), http://www.ociw.edu/ociw/symposia/series/symposium1/ms/\\
gebhardt\_ms.ps.gz
\bibitem[Lawson \& Hanson (1974)]{} Lawson, C. L. \& Hanson, R. J. 
  1974, Solving Least-Squares Problems, Prentice-Hall, Englewood Cliffs, New Jersey
\bibitem[Golub et al.(1999)]{Golub et al.(1999)} Golub, G.
  H., Hansen, P. C., O'Leary, D. P., 1999, SIAM J. Matrix Anal.
  Appl., 21, 185
\bibitem[Hunter \& Qian (1993)]{} Hunter, C., \& Qian, E.~E. 1993, MNRAS, 262, 
  401 (HQ)
\bibitem[Merritt(1996a)]{1996AJ....111.2462M} Merritt, D.\ 1996a, AJ, 111, 
  2462
\bibitem[Merritt(1996b)]{1996AJ....112.1085M} Merritt, D.\ 1996b, AJ, 112, 
  1085 
\bibitem[Merritt(1997)]{1997AJ....114..228M} Merritt, D.\ 1997, AJ, 114, 
  228 
\bibitem[Press et al. 1992]{} Press, W.~H., Teukolsky, S.~A., 
  Vetterling, W.~T., \& Flannery, B.~P. 1992, Numerical Recipes
  (Cambridge: Cambridge University Press)
\bibitem[Richstone \& Tremaine(1984)]{1984ApJ...286...27R} Richstone, 
  D.~O.~\& Tremaine, S.\ 1984, ApJ, 286, 27 
\bibitem[van der Marel et al. 1998]{vdM98} van der Marel, R. P., 
  Cretton, N., de Zeeuw, P. T. \& Rix, H. W. 1998, ApJ, 493, 613 
\bibitem[Verolme \& de Zeeuw(2002)]{2002MNRAS.331..959V} Verolme, E.~K.~\& 
  de Zeeuw, P.~T.\ 2002, MNRAS, 331, 959 
\bibitem[Verolme et al.(2002)]{2002MNRAS.335..517V} Verolme, E.~K.~et al.\ 
  2002, MNRAS, 335, 517
\bibitem[Wahba (1990)]{1990Wahba} Wahba, G. 1990, Spline Models for Observational 
Data (SIAM, Philadelphia)
\bibitem[Valluri, Merritt, Emsellem 2003]{VME03}
  Valluri, M., Merritt, D., Emsellem, E., 2003, ApJ, in press, ApJ preprint doi: 10.1086/380896 
  (astro--ph/0210379, VME03)

\end{thebibliography}
\end{document}